\documentclass[
 aip,
 jmp,
 amsmath,amssymb,
reprint,
]{revtex4-1}

\usepackage{graphicx}
\usepackage{dcolumn}
\usepackage{bm}

\usepackage{docs}
\usepackage{bm}
\usepackage[colorlinks=true,linkcolor=blue]{hyperref}
\expandafter\ifx\csname package@font\endcsname\relax\else
 \expandafter\expandafter
 \expandafter\usepackage
 \expandafter\expandafter
 \expandafter{\csname package@font\endcsname}
\fi
\hypersetup{citecolor = blue}

\begin{document}

\title{Spin-dependent recombination in Czochralski silicon containing oxide precipitates}

\author{V. Lang}
\altaffiliation{Corresponding author. volker.lang@materials.ox.ac.uk.}
\affiliation{Department of Materials, University of Oxford, Oxford OX1 3PH, United Kingdom}

\author{J. D. Murphy}
\affiliation{Department of Materials, University of Oxford, Oxford OX1 3PH, United Kingdom}

\author{R. J. Falster}
\affiliation{Department of Materials, University of Oxford, Oxford OX1 3PH, United Kingdom}
\affiliation{MEMC Electronic Materials Inc., Viale Gherzi 31, 28100 Novara, Italy}

\author{J. J. L. Morton}
\affiliation{Department of Materials, University of Oxford, Oxford OX1 3PH, United Kingdom}
\affiliation{CAESR, Clarendon Laboratory, Department of Physics, University of Oxford, Oxford OX1 3PU, United Kingdom}

\date{\today}

\begin{abstract}
Electrically detected magnetic resonance is used to identify recombination centers in a set of Czochralski grown silicon samples processed to contain strained oxide precipitates with a wide range of densities ($\sim 1 \cdot 10^{9}$ cm$^{-3}$ to $\sim 7 \cdot 10^{10}$ cm$^{-3}$). Measurements reveal that photo-excited charge carriers recombine through  P$_{\text{b0}}$ and P$_{\text{b1}}$ dangling bonds and comparison to precipitate-free material indicates that these are present at both the sample surface and the oxide precipitates. The electronic recombination rates vary approximately linearly with precipitate density. Additional resonance lines arising from iron-boron and interstitial iron are observed and discussed. Our observations are inconsistent with bolometric heating and interpreted in terms of spin-dependent recombination. Electrically detected magnetic resonance is thus a very powerful and sensitive spectroscopic technique to selectively probe recombination centers in modern photovoltaic device materials.
\end{abstract}

\keywords{silicon, EDMR, spin-dependent recombination, oxide precipitate, lifetime}

\maketitle

\section{Introduction}
Oxygen is an important impurity in silicon, being present in concentrations of $\sim 10^{18}$ cm$^{-3}$ in Czochralski silicon (Cz-Si), which is used for the vast majority of integrated circuits (ICs) and $\sim 40\%$ of solar cells. It is also present in lower, but still significant, concentrations (of order $10^{17}$ cm$^{-3}$) in cast multicrystalline silicon (mc-Si) used equally if not more widely for modern silicon photovoltaics. The presence of oxygen has substantial beneficial as well as detrimental effects on silicon's material properties and so has been the subject of much research (see \cite{borghesi1,newman1}). Perhaps most importantly, oxide precipitates (OPs) can be intentionally created in inactive regions of wafers to act as sinks for detrimental metallic impurities in a process known as \textit{internal gettering} \cite{tytan1,smmyers1,dgilles1}. Oxygen can also improve high temperature mechanical strength by atomic decoration of dislocations \cite{agiannattasio1} and by precipitation in the bulk \cite{kjurkschat1}. Unfortunately, oxygen-containing defects in various guises also act as recombination centers, including thermal donor defects \cite{fuller1}, boron-oxygen complexes \cite{hfischer1,bothe1}, and OPs \cite{vanhellemont1,jdmurphy1}. OPs can also form unintentionally in mc-Si during ingot cooling \cite{bothe2}, and can limit the efficiency of modern silicon photovoltaic devices \cite{chen1}. They undergo a morphological transformation during growth \cite{falster1,bergholz1}, which is known to have implications for both, internal gettering \cite{falster1} and recombination of minority carriers \cite{jdmurphy1}. The precipitates initially exist in an unstrained state (sometimes referred to as ``ninja particles''), but after a certain threshold growth time (dependent on the density of nucleation sites) they change morphology into a strained state, which coincides with the transition from ineffective to totally effective gettering \cite{falster1}. They then continue to grow in size and eventually begin to become surrounded by complex dislocation structures and even stacking faults \cite{falster1}. Recombination at OPs has recently been found to depend upon their strain state and whether the precipitates are surrounded by other extended defects \cite{jdmurphy1}. Interestingly, the rate of recombination in samples dominated by strained precipitates is dependent upon precipitate density, rather than size \cite{jdmurphy1}. As the number of corners of the strained platelets is invariant with size, these discontinuities have been suggested to play a role in the recombination process \cite{jdmurphy1}. The photoconductance methods \cite{sinton1} used in this previous study \cite{jdmurphy1} do, however, not allow the microscopic nature of the specific defect(s) responsible for recombination to be clearly determined. \\

In this paper we present the results of experiments using electrically detected magnetic resonance (EDMR), which aim to better understand the recombination mechanism associated with OPs. EDMR is a sensitive spectroscopic technique providing a much higher sensitivity than conventional electron paramagnetic resonance (EPR) for bulk samples, which has been used extensively to study defects and impurities in silicon \cite{hhwoodbury1,yhlee1,ehpoindexter1,wecarlos1,klbrower1,astesmans1}. The sensitivity of EDMR is typically $\sim 10^6$ times higher than for conventional EPR, and has been demonstrated to approach the few to single spin regime for nanodevices with an optimised sample geometry \cite{drmccamey1,cboehme1}. In EDMR, the sample is placed in a static magnetic field and irradiated in a microwave cavity. The EPR-induced change in spin population is detected through the (resonant) change of the device conductivity. Hence, only electrically active defects involved in electron transport, such as recombination centers, are observed in EDMR. As opposed to EPR, parasitic signals, which do not determine electron transport, are consequently not observed. This is why EDMR has been particularly successful in the spectroscopic characterization of e.g. electron transport in silicon field-effect transistors \cite{cclo1} as well as defects and impurities in modern silicon photovoltaic devices \cite{mstutzmann1}. Magnetic resonance techniques have also been used previously to study recombination at OPs \cite{tmchedlidze1,mkoizuka1}, showing, for example, that silicon dangling bonds at the silicon-silicon dioxide interface (Si/SiO$_2$) play a major role in recombination at OPs \cite{mkoizuka1}. In our work we study materials possessing a wide range of strained precipitate densities, which we relate to spin-dependent recombination times extracted from the EDMR data. We also study intentionally iron-contaminated samples to improve understanding of the role of interstitial iron (Fe) in the recombination process.
\begin{figure}[t]
\includegraphics[scale=0.88]{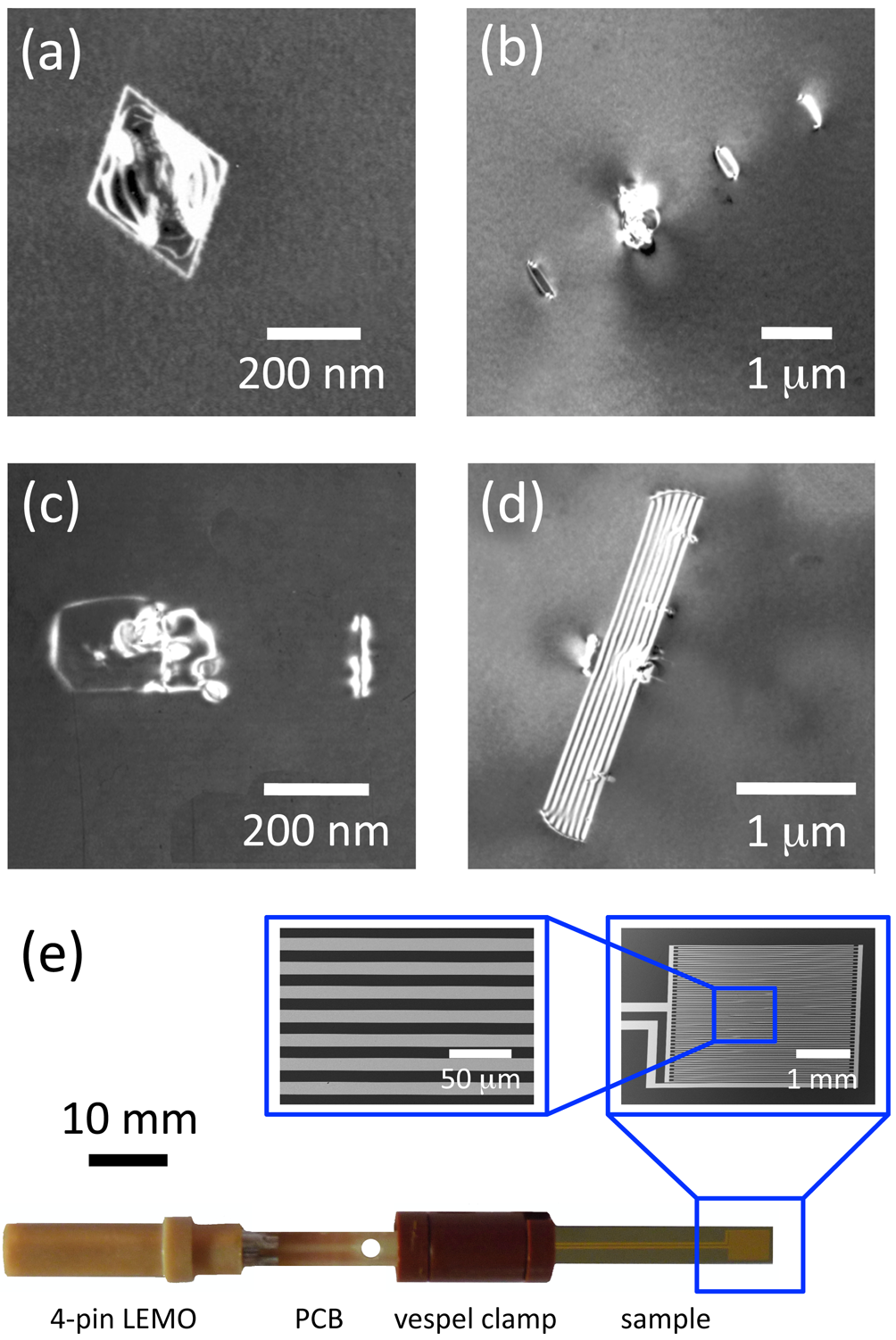}
\caption{(Color online) (a-d) Bright-field TEM micrographs of typical strained OPs. The strain contrast in (a) reveals their square platelet-like shape. (b) and (c) show a platelet surrounded by dislocation loops, the OP shown in (d) is surrounded by a stacking fault. (e) Optical image of the overall sample arrangement (bottom). The sample is connected to a printed circuit board equipped with a 4-pin LEMO 00-Series connector with a miniature vespel clamp. SEM images of the interdigitated contact geometry are shown at a magnification of $220\times$ (top, right) and $2,000\times$ (top, left).}
\label{fig1}
\end{figure}

\begin{table}[b]
\begin{ruledtabular}
\begin{tabular}{lrrr}
& \textbf{nucleation} & \textbf{growth} & \\
\vspace{0.8mm}
\textbf{initial [O]} & \textbf{time (h)} & \textbf{time (h)} & \textbf{[OP] ($10^{10}\:$cm$^{-3}$)}\\
\hline
low & 6 & 8 & $0.08 \pm 0.002$ \\
low & 8 & 8 & $0.15 \pm 0.031$ \\
low & 6 & 8 & $0.22 \pm 0.017$ \\
low & 8 & 16 & $0.28 \pm 0.033$ \\
low & 16 & 8 & $0.64 \pm 0.033$ \\
low & 16 & 8 & $1.29 \pm 0.217$ \\
low & 16 & 16 & $2.27 \pm 0.359$ \\
low & 32 & 8 & $4.55 \pm 0.768$ \\
low & 32 & 16 & $6.98 \pm 0.133$ \\
high & 8 & 16 &$ 7.04 \pm 0.789$
\end{tabular}
\caption{Details of the samples processed to contain different concentrations [OP] of strained OPs. }
\label{tab1}
\end{ruledtabular}
\end{table}

\section{Experimental methods}
\subsection{Growth of oxide precipitates}
Samples were prepared from $150\:$mm diameter (100)-orientated high-purity p-type Cz-Si wafers doped with boron (B). The carbon content of all wafers was below the detection limit of $\sim 5 \cdot 10^{15}$ cm$^{-3}$. ``Low" oxygen concentration wafers had [O] $= 7.8 \cdot 10^{17}$ cm$^{-3}$ and [B] $= (1.3 \pm 0.2) \cdot 10^{15}$ cm$^{-3}$, ``high" oxygen concentration wafers [O] $= 9.3 \cdot 10^{17}$ cm$^{-3}$ and [B] $= (1.1 \pm 0.1) \cdot 10^{15}$ cm$^{-3}$. The oxygen concentrations are stated to the DIN50438/I (1995) standard. OPs were created by carefully controlled heat treatments under ultra-clean conditions described in detail elsewhere \cite{kfkelton1,jdmurphy1}. In summary, the precipitation treatment used has four-stages comprising a homogenization ($15$min at $1000\:^{\circ}$C), nucleation ($6-32$h at $650\:^{\circ}$C), drift ($4$h at $800\:^{\circ}$C), and growth ($8-16$h at $1000\:^{\circ}$C) anneal. Transmission electron microscopy (TEM) micrographs of typical strained OPs are shown in Fig.~\ref{fig1} (a-d). Fig.~\ref{fig1} (a) reveals that strained OPs are platelets \cite{bergholz1}. After sufficient growth time their strain field promotes the formation of crystal defects, such as dislocation loops (Fig.~\ref{fig1} (b) and (c)) and stacking faults as shown in Fig.~\ref{fig1} (d). The concentration of strained OPs was varied by using different durations for the nucleation and growth anneals. Precipitate densities were measured by using a Schimmel etch on cross-sections of pieces from the wafer from which the samples were taken. The measured OP densities as well as the particular growth conditions of each sample are given in Table~\ref{tab1}. A precipitate-free sample of Cz-Si with a boron concentration of [B] $= 8.0 \cdot 10^{14}$ cm$^{-3}$ was used for control purposes.

\subsection{Iron: Concentration measurement and contamination}
Bulk iron concentrations were determined in samples subjected to identical precipitation treatments using photoconductance methods \cite{sinton1}, which are well-established and based on photodissociation of iron-boron pairs \cite{zoth1,jdmurphy1,jdmurphy2}. In all cases the bulk iron concentration was below $3 \cdot 10^{11}$ cm$^{-3}$\ \cite{jdmurphy1}, confirming the precipitation treatments had been performed in clean conditions. A sample with a strained precipitate density of $0.15 \cdot 10^{10}\:$cm$^{-3}$ was intentionally post-contaminated by rubbing its backside with a piece of iron ($99.9\%$ purity) followed by subsequent annealing at $775\:^{\circ}$C for $23$h. The bulk iron concentration in an identically processed sample from the same precipitate-containing wafer was measured to be $1.2 \cdot 10^{12}$ cm$^{-3}$, compared to $2.5 \cdot 10^{12}$ cm$^{-3}$ in an identically-processed precipitate-free sample \cite{jdmurphy2}. The lower bulk iron concentration in the precipitate-containing sample strongly indicates that $\sim 1 \cdot 10^{12}$ cm$^{-3}$ of the bulk iron is gettered to the OPs and associated defects.

\begin{figure}[t]
\includegraphics[scale=0.9]{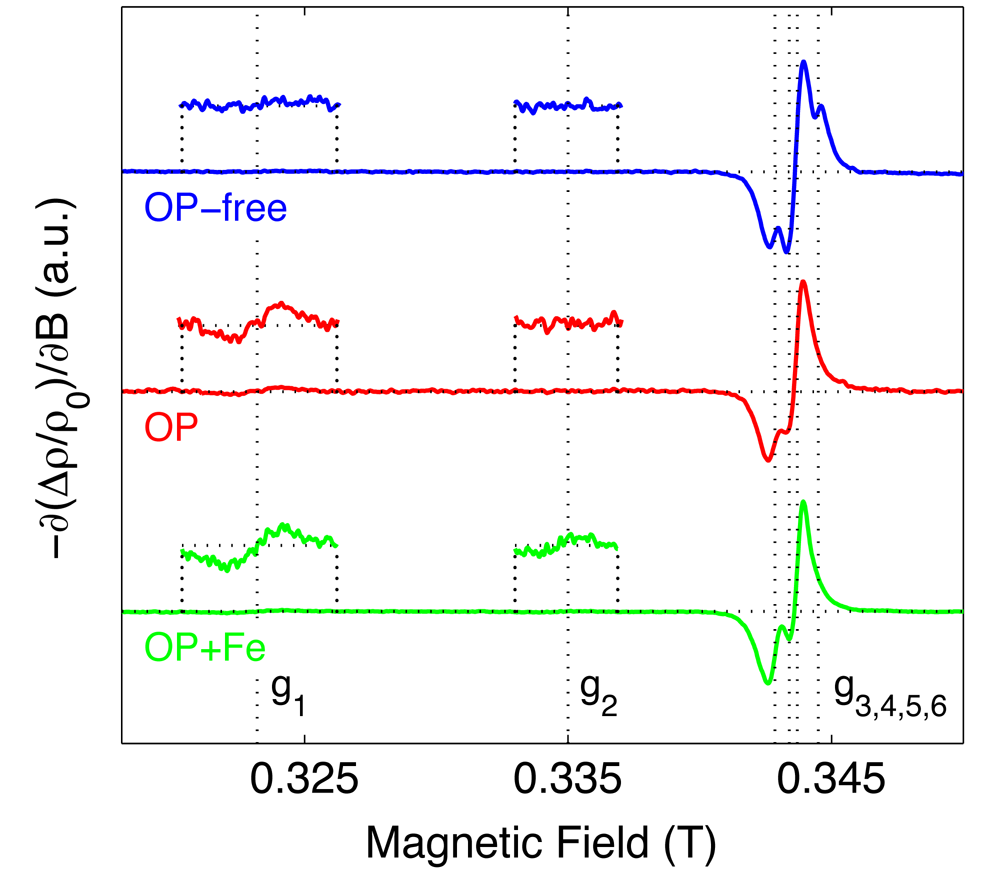}
\caption{(Color online) Comparison of normalized EDMR spectra from a precipitate-free (top, blue), a not intentionally contaminated (middle, red), and an iron-contaminated sample (bottom, green). The middle and bottom samples have [OP] $= 0.15 \cdot 10^{10}\:$cm$^{-3}$. Larger scale plots are shown for low intensity peaks and vertical lines (dotted) indicate the magnetic field values of the different Land\'e $g$ factors $g_{1,2,...,6}$. All spectra are offset for clarity.}
\label{fig2}
\end{figure}

\subsection{EDMR experiments}
Interdigitated chromium/gold (10/30$\:$nm) thin films were defined by electron-beam lithography on a JEOL JBX-5500FS system and used to electrically contact the active area of the device with a minimal resistivity and without perturbing the microwave field in the cavity. All samples were diced into small chips of $2 \times 20\:$mm$^2$ in size and connected to a printed circuit board with a miniature vespel clamp as shown in Fig.~\ref{fig1} (e). The maximum diameter of the overall arrangement was chosen so that it can be inserted into a glass tube with an outer diameter of $5\:$mm to protect the sample against mechanical damage during its insertion into the cavity and vibration during the measurement. EDMR was performed with a modified Bruker ESP380-1010 pulsed X-band EPR spectrometer and an Oxford Instruments CF-935 helium-gas flow cryostat in combination with an Oxford Instruments ITC-503S temperature controller. The microwave excitation was generated with an Agilent Technologies E8267D PSG vector signal generator and applied by a Bruker ER4118X-MD5-W1 X-band dielectric ring resonator operating at a DC magnetic field $B_0 \sim 0.35\:$T. The magnetic field was aligned perpendicular to the growth direction of the sample, i.e.\ $\mathbf{B_0} \parallel [100]$ (unless otherwise indicated). Magnetic field modulation was applied with a HP 33120A function generator in order to enhance the signal-to-noise ratio. This lock-in technique results in the EDMR signal appearing as the first derivative of the sample resistivity with respect to magnetic field, i.e. $\partial(\Delta\rho/\rho_0)/\partial B$, where $\rho_0$ denotes the sample resistivity in thermal equilibrium\cite{footnote1}. The magnetic field as well as the modulation amplitude was calibrated with a 2,2-diphenyl-1-picrylhydrazyl (DPPH) reference sample and amounted to $0.1\:$mT with a modulation frequency of $5.02\:$kHz (unless otherwise indicated). A battery-powered variable resistor network was used to apply a constant current to the sample of typically $I = 20 - 100\: \mu$A. The sample was placed  under constant illumination with a Schott KL1500 150W halogen cold light source and the resonant change of the voltage drop across the sample was detected via a FEMTO DLPVA-100-F-D variable gain low-noise differential voltage amplifier and a SR830 lock-in amplifier. All measurements were carried out at a temperature $T = 60\:$K, which was found to be the optimal in the signal-to-noise ratio.

\begin{figure}[t]
\includegraphics[scale=0.9]{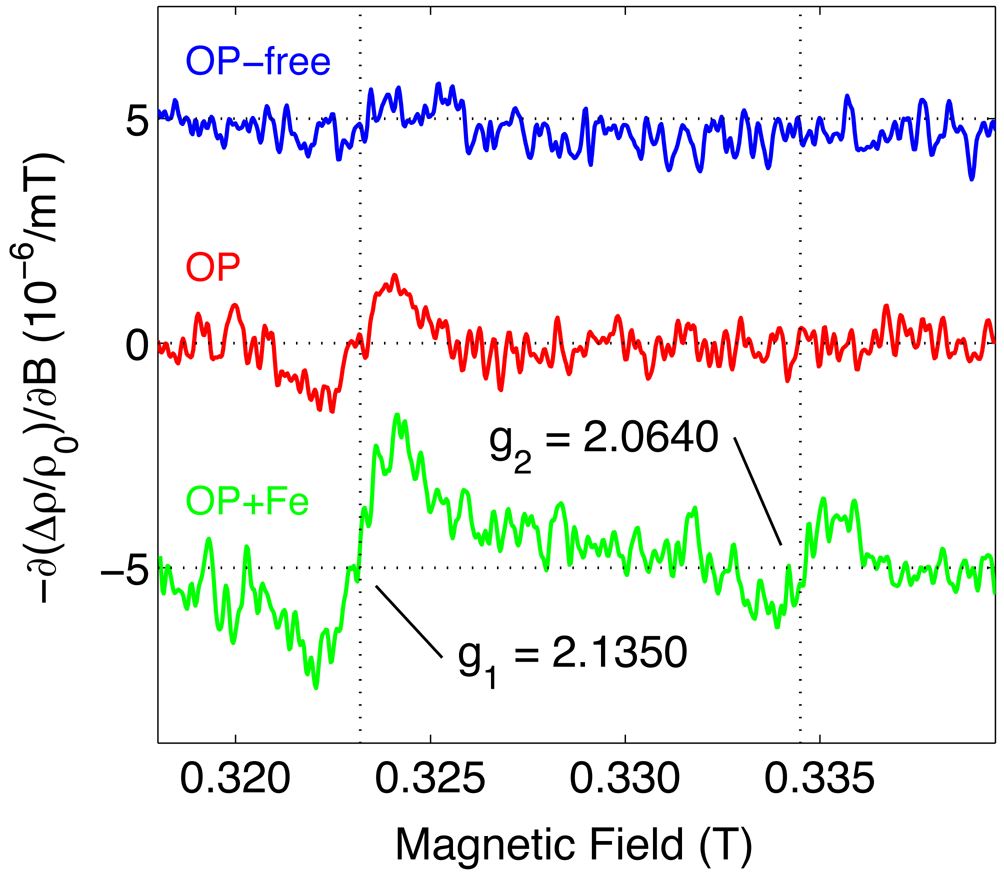}
\caption{(Color online) Comparison of the low-field resonances $g_1$ and $g_2$ for the same set of samples shown in Fig.~\ref{fig2}. Vertical lines (dotted) indicate the magnetic field values of $g_1$ and $g_2$, respectively. All traces are offset for clarity.}
\label{fig3}
\end{figure}

\section{Results}
Typical EDMR spectra obtained from (i) a precipitate-free, (ii) a not intentionally contaminated sample with OPs, and (iii) an iron-contaminated sample with OPs ([OP] $= 0.15 \cdot 10^{10}\:$cm$^{-3}$) are shown in Fig.~\ref{fig2}. In each case, the main resonance feature is a superposition of multiple resonance lines. The Land\'e $g$ factor of each resonance is indicated and resonances are numbered from left to right according to the order in which they are discussed below. The Land\'{e} $g$ factors identified are stated in Table~\ref{tab2}.

\begin{figure}[t]
\includegraphics[scale=0.9]{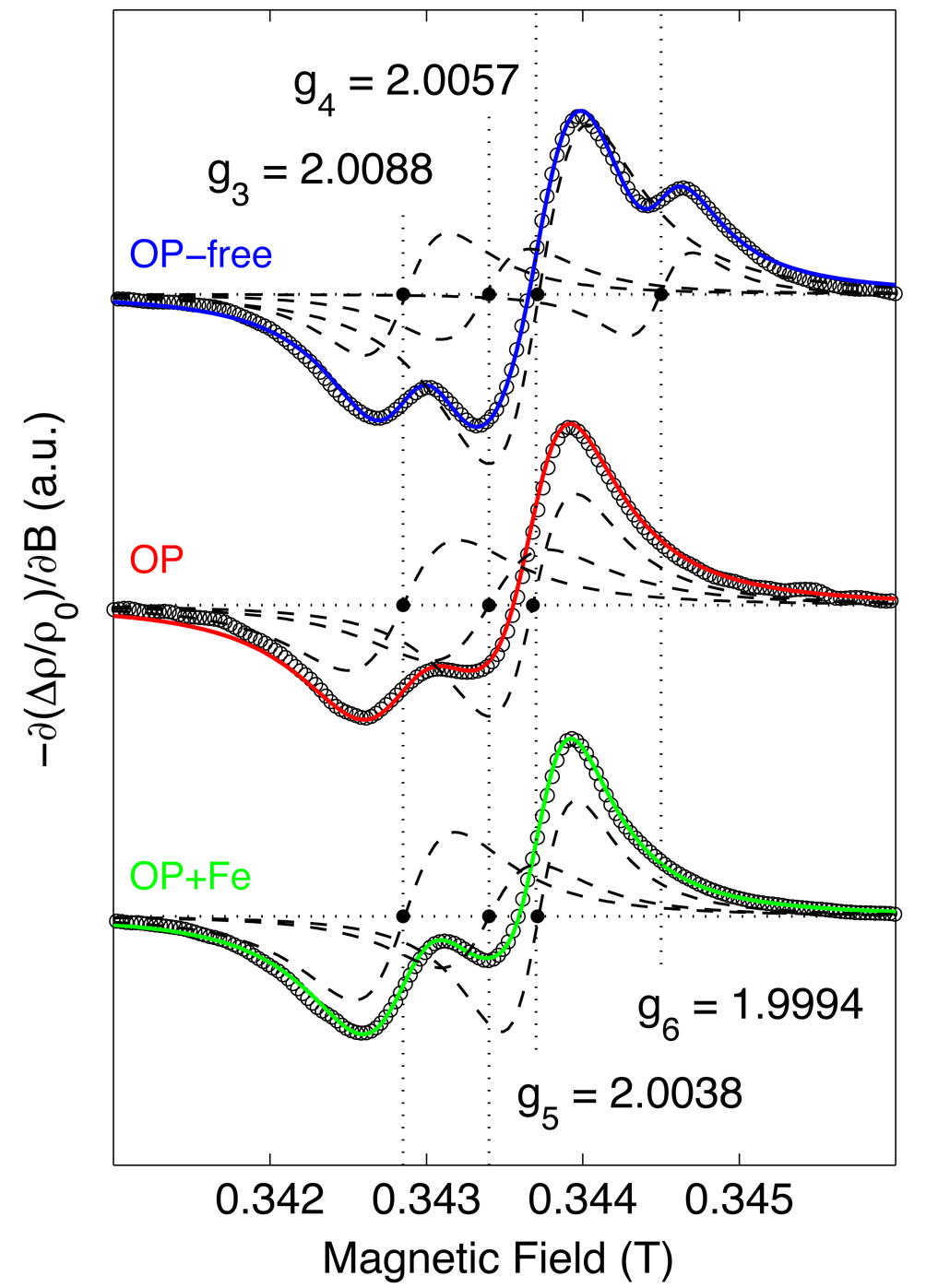}
\caption{(Color online) Comparison of the main resonances for the same samples as in Fig.~\ref{fig2}. Solid lines represent numerical best fits of the data ($\circ$) to a superposition of four (top) and three (middle and bottom) Lorentzian derivative lines (dashed, black), respectively. Traces are offset for clarity and solid points ($\bullet$) highlight the magnetic field values of the observed Land\'e $g$ factors $g_{3,4,5,6}$.}
\label{fig4}
\end{figure}

\subsection{FeB pair and interstitial Fe}
A comparison of the two lower-field features of the EDMR spectra is shown in Fig.~\ref{fig3}. Each spectrum reveals a resonance at $g_1 = 2.1350$ with the signal intensity being weakest in the precipitate-free and strongest in the iron-contaminated sample. Interstitial Fe is well known to bind to substitutional B to form FeB pairs, so we assign $g_1$ to the FeB pair in accordance with the literature\cite{aaistratov1}.  Another weak signal is observed in the iron-contaminated sample at $g_2 = 2.0640$ and assigned to interstitial Fe in good agreement with previous EDMR experiments \cite{wgehlhoff1}. The co-existence of the FeB and Fe defects is to be expected as the FeB pair dissociates under illumination with white light \cite{zoth1,jdmurphy2} and the intensity of the illumination used in our experiments is unlikely to be sufficient for complete dissociation.

\begin{table}[b]
\begin{ruledtabular}
\begin{tabular}{lll}
\vspace{0.8mm}
\textbf{our experiment} & \textbf{literature} & \textbf{nature of defect} \\
\hline
$g_1 = 2.1350$ & 2.1345\cite{aaistratov1} & FeB pair \\
$g_2 = 2.0640$ & 2.0690\cite{hhwoodbury1,yhlee1,ehpoindexter1,wgehlhoff2,wgehlhoff1} & interstitial Fe \\
$g_3 = 2.0088$ & 2.0086\cite{wecarlos1,klbrower1,mkoizuka1,astesmans1} & P$_{\text{b0}}$ dangling bond \\
$g_4 = 2.0057$ & 2.0056\cite{mkoizuka1,astesmans1} & P$_{\text{b1}}$ dangling bond\\
$g_5 = 2.0038$ & 2.0037\cite{wecarlos1,klbrower1,mkoizuka1,astesmans1} & P$_{\text{b0}}$ dangling bond \\
$g_6 = 1.9994$ & & unknown
\end{tabular}
\caption{Comparison between the most important Land\'e $g$ factors in Cz-Si with OPs observed in our experiments with their corresponding values reported in literature. The microscopic nature of each resonance is indicated in the last column.}
\label{tab2}
\end{ruledtabular}
\end{table}

\begin{figure}[t]
\includegraphics[scale=0.94]{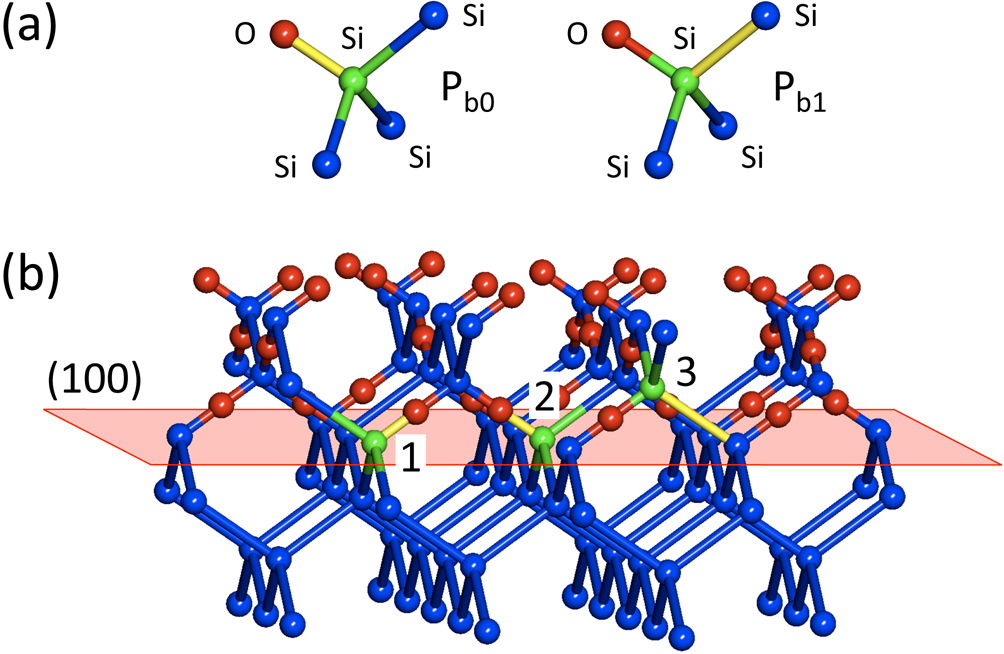}
\caption{(Color online) (a) Chemical nature of P$_{\text{b0}}$ and P$_{\text{b1}}$ dangling bonds. Each dangling bond defect consists of three components: a central Si atom (green) with a dangling orbital (yellow) and an O atom (red). The Si atom of the P$_{\text{b0}}$ (P$_{\text{b1}}$) dangling bond is back-bonded to three (two) other Si atoms (blue) of the surrounding matrix. (b) shows the crystallographic nature of P$_{\text{b0}}$ (1,2) and P$_{\text{b1}}$ dangling bonds (3) in an exemplary section of the (100) Si/SiO$_2$ interface. The P$_{\text{b0}}$ center has two different crystallographic orientations (1,2) relative to the interface plane (red, shaded).}
\label{fig5}
\end{figure}

\subsection{P$_{\text{b0}}$ and P$_{\text{b1}}$ dangling bonds}
A detailed comparison of the higher-field features of the spectra (close to $g=2.0$) is shown in Fig.~\ref{fig4}. The spectra of the samples containing OPs are fit best by a superposition of three Lorentzian derivative lines with the Land\'e $g$ factors $g_{3,4,5}$, whereas the spectrum of the precipitate-free sample is fit best by a superposition of four Lorentzian derivative lines with $g_{3,4,5,6}$. In other words, $g_{3,4,5}$ were observed in all samples with and without  OPs, whereas $g_6$ was only observed in the precipitate-free sample. It is therefore possible that the paramagnetic center associated with $g_6 = 1.9994$ is annealed out during the high temperature growth of the OPs. $g_5 = 2.0038$ and $g_3 = 2.0088$ correspond to the two crystallographic orientations of P$_{\text{b0}}$ dangling bonds at (100) Si/SiO$_2$ interfaces and are in good agreement with literature \cite{wecarlos1,klbrower1,mkoizuka1,astesmans1} (see Table~\ref{tab2}). $g_4 = 2.0057$ is interpreted in terms of P$_{\text{b1}}$ dangling bonds\cite{astesmans1}. 
\begin{figure}[t]
\includegraphics[scale=0.9]{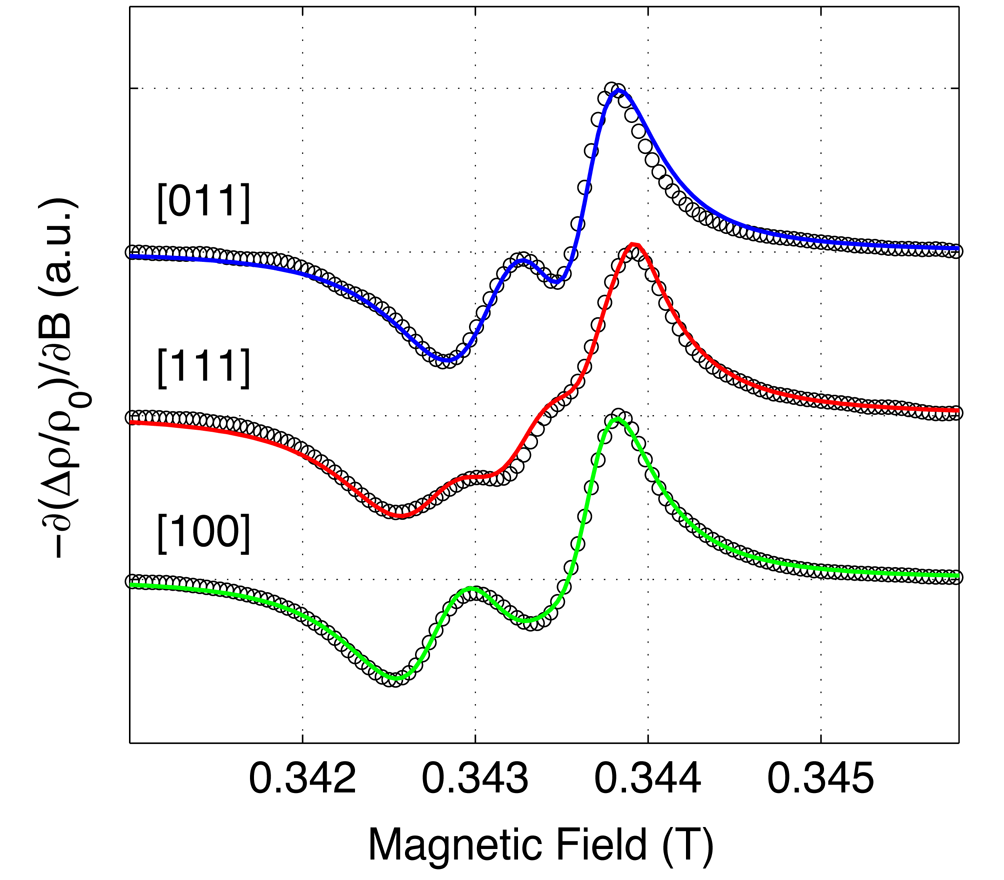}
\caption{(Color online) Angular dependence of the main resonances of the iron-contaminated sample with OPs for $\mathbf{B_0}\parallel [011]$ (top), $[111]$ (middle), and $[100]$ (bottom). Solid lines represent numerical best fits of the data ($\circ$) to a superposition of three, four, and two Lorentzian derivative lines, respectively. All spectra are normalized and offset for clarity.}
\label{fig6}
\end{figure}
In order to verify this interpretation we investigate the angular dependence of $g_{3,4,5}$, which allows us to identify the nature of these defects unambiguously owing to their inherent crystallographic symmetry. The crystallographic nature of both defects is shown schematically in Fig.~\ref{fig5}. The P$_{\text{b0}}$ center $\bullet\:$Si $\equiv$ Si$_3$ is formed by a trivalent silicon atom back-bonded to three other silicon atoms, whereas the P$_{\text{b1}}$ center is associated with \linebreak $\bullet\:$Si $\equiv$ Si$_2$O, i.e.\ a partially oxidized silicon atom with a dangling orbital. Fig.~\ref{fig6} shows the EDMR spectrum of the iron-contaminated sample with OPs for different orientations of the sample surface with respect to the magnetic field $\mathbf{B_0}$. The different spectra for $\mathbf{B_0} \parallel [011], [111],$ and $[100]$ are numerically best fit to a superposition of three (top), four (middle), and two (bottom) Lorentzian derivative lines. The resulting Land\'e $g$ factors are compiled in Table~\ref{tab3}. For an axial (trigonal) symmetry, the shift as well as the disappearance and reappearance of the different resonances as a function of the rotation angle $\theta$ between $\mathbf{B_0}$ and the $[100]$ sample normal is described by
\begin{equation} \label{eqn1}
g(\theta) = \sqrt{(g_{||} \cos \theta)^2 + (g_{\perp} \sin \theta)^2}
\end{equation}
where $g_{||}$ and $g_{\perp}$ denote the principle $g$ values of the P$_{\text{b0}}$ and P$_{\text{b1}}$ dangling bonds, respectively. A comparison between the experimentally observed and theoretically predicted values is shown in Table~\ref{tab3}. The good agreement of the experimental and theoretical values obtained from Eqn. (\ref{eqn1}) supports our interpretation of $g_{3,5}$ in terms of P$_{\text{b0}}$, and $g_4$ in terms of P$_{\text{b1}}$ dangling bonds. Both defects have been associated with oxygen precipitation in a series of EPR experiments by Koizuka \textit{et al.}~before \cite{mkoizuka1} without ruling out the possibility of recombination through dangling bonds at the sample surface. Mchedlidze \textit{et al.} carried out a similar study but did not observe any EDMR signal in Cz-Si with OPs \cite{tmchedlidze1}. This led them to propose that only iron-decorated OPs can act as recombination centers, which is in contrast to the experiments of Koizuka \textit{et al.}~as well as our own observations.

\begin{table}[b]
\begin{ruledtabular}
\begin{tabular}{lll}
\vspace{0.8mm}
\textbf{direction of $\mathbf{B_0}$} & \textbf{$g$ value of P$_{\text{b0}}$} & \textbf{$g$ value of P$_{\text{b1}}$} \\
\hline
$[011]$ & 2.0088  (2.0086) & 2.0054 (2.0056) \\
\vspace{1mm}
               & 2.0034 (2.0037) & \\
$[111]$ & 2.0078 (2.0078) & 2.0046 (2.0045) \\
\vspace{1mm}
                 & 2.0013 (2.0013) & 2.0022 (2.0021) \\
$[100]$ & 2.0063 (2.0063) & 2.0031 (2.0031) \\
\end{tabular}
\caption{Angular dependence of the Land\'e $g$ factors associated with P$_{\text{b0}}$ and P$_{\text{b1}}$ dangling bonds. Theoretical values obtained from (\ref{eqn1}) with the principal values from Ref. \cite{astesmans1} are parenthesized.}
\label{tab3}
\end{ruledtabular}
\end{table}

\subsection{Influence of OP density on recombination time}
In order to identify and better understand the underlying process giving rise to the EDMR signals, the electronic recombination times associated with recombination through P$_{\text{b0}}$ and P$_{\text{b1}}$ dangling bonds (i.e.\ $g_{3,5}$ and $g_4$, respectively) have been measured in Cz-Si samples with different concentrations of OPs [OP]$=(0.15 - 7.04) \cdot 10^{10}\:$cm$^{-3}$ (see Table~\ref{tab1}). Determination of this lifetime assumes that the photocurrent through the device is determined by the time-dependent density of photo-excited electrons $n(t)$, which is described by the rate equation
\begin{equation} \label{diffeqn}
\dot n(t) = g-nr = g - nr_0 (1+ae^{i \omega_{mod} t}) \; ,
\end{equation}
where $g$ is the generation rate, and $r = r_0 (1+ae^{i \omega_{mod} t})$ the recombination rate modulated around its average value $r_0$ with modulation frequency $\omega_{mod}$ and modulation amplitude $a$. Equation \ref{diffeqn} can be solved by making a first-order approximation, i.e. $\mathcal{O}(e^{2 i\omega_{mod}t}) \approx 0$, with the Ansatz $n(t) = n_0 + n_1 e^{i \omega_{mod} t}$. The solution then reads
\begin{equation}
n(t) = n_0 \left(1- a \frac{1-i \omega_{mod} \tau}{1+\omega_{mod}^2 \tau^2} e^{i \omega_{mod} t} \right) \; ,
\end{equation}
where $\tau := 1/r_0$ is the electronic recombination time associated with the particular recombination center, and $n_0 = g\tau$ the average carrier density. The recombination time $\tau$ can be determined by measuring the absolute value of the EDMR signal amplitude $|\Delta\rho|$ as a function of modulation frequency and fitting it to
\begin{equation} \label{solution}
\lvert \Delta\rho \rvert = \bigg\lvert\frac{1-i\omega_{mod}\tau}{1+\omega_{mod}^2 \tau^2} \Delta\rho_0\bigg\rvert  = \frac{\lvert \Delta\rho_0 \rvert}{\sqrt{1+\omega_{mod}^2 \tau^2}} \; .
\end{equation}
The EDMR signal approaches its maximum value $|\Delta\rho_0|$ for $\omega_{mod} < 1/\tau$, and decreases proportionally to $1/\omega_{mod}\tau$ for $\omega_{mod} > 1/\tau$. This technique has been successfully applied previously to study e.g. spin-dependent trapping at trivalent silicon centers in silicon bicrystals\cite{pmlenahan1} as well as spin-dependent recombination at the silicon surface\cite{djlepine1} and P$_{\text{b0}}$ dangling bonds\cite{pmlenahan2}. We note that the modulation frequency has to be slow compared to the spin-lattice relaxation time $T_1$ of the spins in order to comply with the slow-adiabatic passage condition and avoid distortion of the EDMR signal due to passage effects\cite{mweger1}, which have not been observed in our experiments.
\begin{figure}[t]
\includegraphics[scale=0.9]{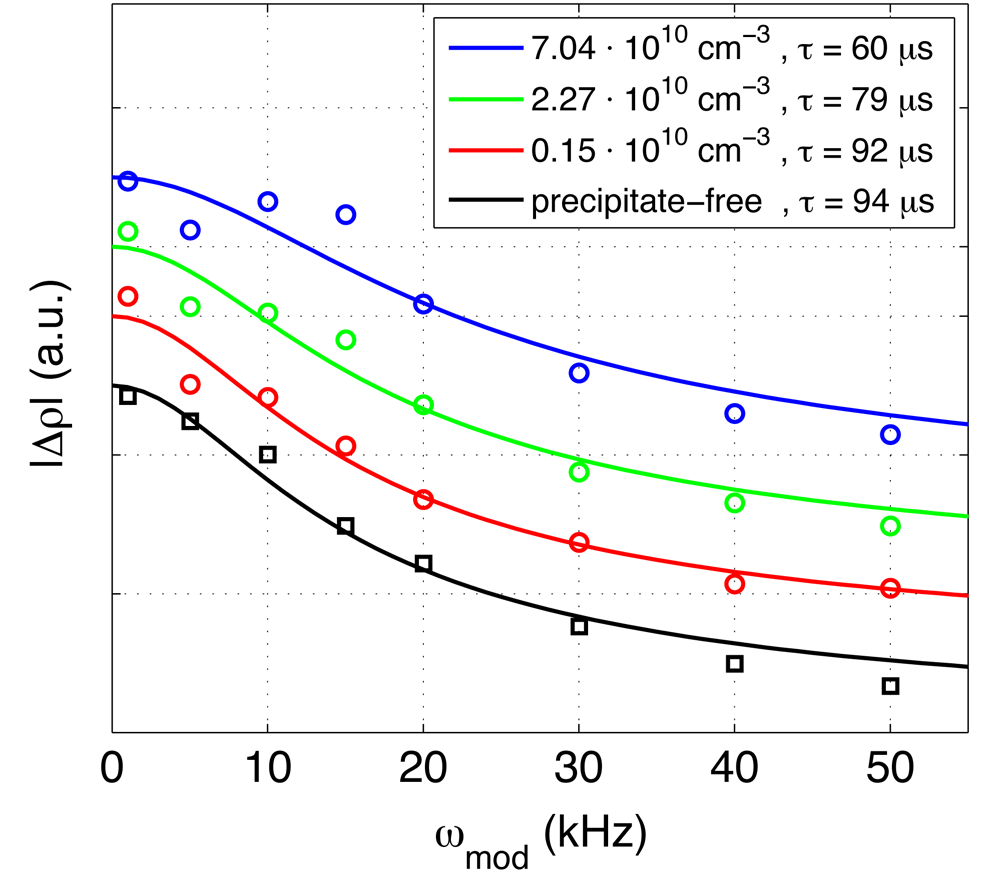}
\caption{(Color online) Absolute EDMR signal amplitude $\lvert\Delta\rho\rvert$ as a function of modulation frequency $\omega_{mod}$ for the precipitate-free sample ($\Box$, black) and three exemplary samples with a low ($\circ$, red), medium ($\circ$, green), and high concentration of OPs ($\circ$, blue). Numerical best fits (solid lines) of the data to Eqn.~(\ref{solution}) are shown. The recombination times are indicated and decrease with an increasing concentration of OPs. All traces are offset for clarity.}
\label{fig7}
\end{figure}

\begin{figure}[b]
\includegraphics[scale=0.9]{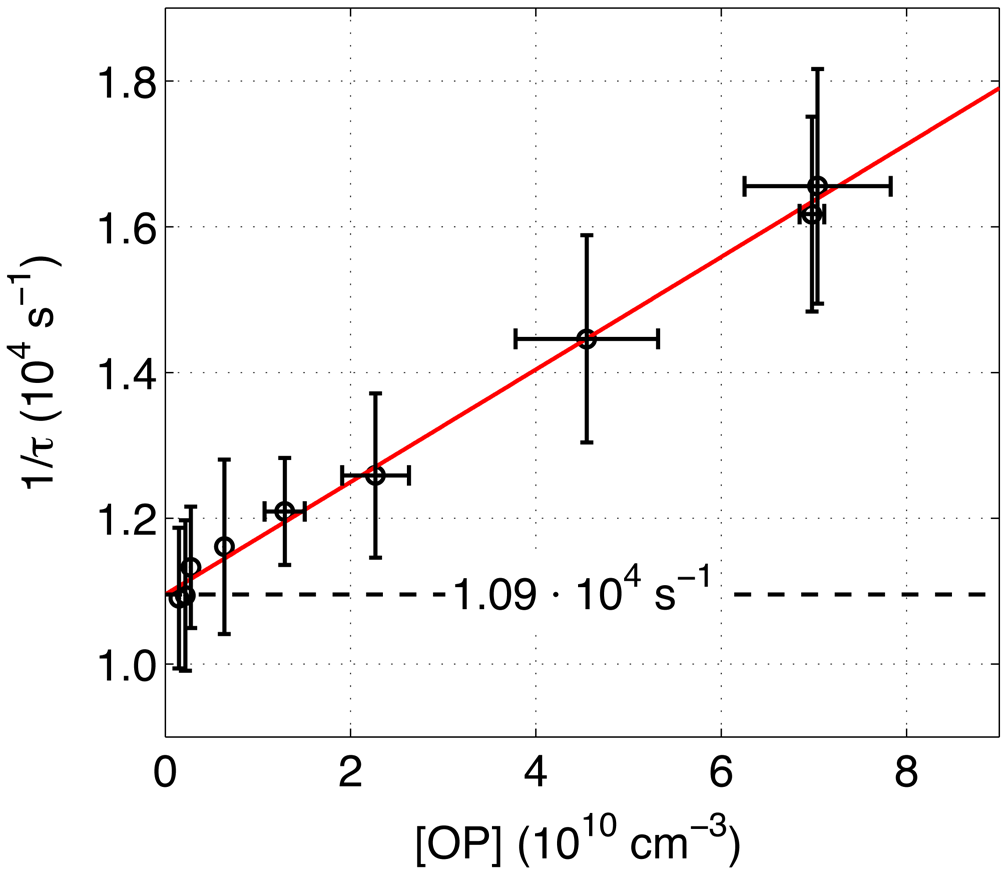}
\caption{(Color online) Reciprocal carrier recombination time $1/\tau$ as a function of strained OP concentration [OP]. The linear fit (solid line, red) to the data ($\circ$) yields a slope of $\gamma = 7.72 \cdot 10^{-8}\:$cm$^3$s$^{-1}$ and a y-axis intercept of $\delta = 1.09 \cdot 10^{4}\:$s$^{-1}$ (dashed line, black).}
\label{fig8}
\end{figure}

We employed this technique to measure the electronic recombination time of P$_{\text{b0}}$ and P$_{\text{b1}}$ dangling bonds in Cz-Si without and with OPs with different concentrations \cite{footnote2}. Our results are shown in Fig.~\ref{fig7} for a set of four exemplary samples using a modulation frequency of $\omega_{mod} \leq 50$~kHz, well below the expected cutoff frequency of our experimental setup $\omega \sim 350$~kHz. The electronic recombination time decreases with an increasing concentration of OPs, and has a maximum value $\tau = 94\:\mu$s in the precipitate-free sample. Similar values have been observed for P$_{\text{b0}}$ centers at the (100) Si/SiO$_2$ surface with this technique before\cite{pmlenahan1}. Previous photoconductance measurements on surface passivated Cz-Si with OPs have shown that the reciprocal of the recombination time $1/\tau$ increases with increasing OP concentration [OP] linearly according to $1/\tau = \gamma$[OP], where $\gamma$ denotes the capture coefficient \cite{jdmurphy1}. Our measurements reveal a similar relationship but with a non-zero y-axis intercept as shown in Fig.~\ref{fig8}. Our data is best fit to $1/\tau = \gamma$[OP]$+\delta$ with $\gamma = 7.72\cdot 10^{-8}\:$cm$^3$s$^{-1}$ and $\delta = 1.09 \cdot 10^{4}\:$s$^{-1}$, which corresponds to a recombination time $\tau = 1/\delta = 92\:\mu$s. The recombination time of dangling bonds in the iron-contaminated sample was measured to $\tau(\text{P}_{\text{b0,b1}}) = 92\:\mu$s, the lifetime of the iron-boron pair to $\tau(\text{FeB}) = 64\:\mu$s, which is significantly shorter than most of the recombination times observed in our samples.

\section{Discussion}
The sign of each EDMR signal shown in Fig.~\ref{fig2} was determined by measuring the DC change in sample resistivity on and off resonance. This measurement confirms that the sample resistivity increases upon resonance, i.e.\ $\Delta\rho/\rho_0 > 0$, which corresponds to a resonant decrease of the photocurrent. At the same time, the sample resistivity has been observed to increase with decreasing temperature for all samples, i.e. $\partial\rho_0/\partial T < 0$. Hence, bolometric heating can be ruled out as the predominant EDMR mechanism as that would produce an EDMR signal, which follows the sign of $\partial\rho_0/\partial T$. Furthermore, a resonance line associated with conduction band electrons has not been observed in our experiments. We therefore interpret our results in terms of spin-dependent recombination \cite{djlepine1} of photo-excited electron-hole pairs from the Urbach tails of the conduction band \cite{furbach1} through the recombination centers compiled in Table~\ref{tab2}. The two lower-field features of the EDMR spectra shown in Fig.~\ref{fig3} have been explained in terms of the FeB pair and the Fe center. The observation of the FeB pair is particularly interesting as this impurity has, to our best knowledge, not been observed in EDMR before. Owing to its comparatively short electronic recombination time $\tau = 64\: \mu$s, our results demonstrate that the FeB pair is a very active recombination center and contributes to spin-dependent recombination in Cz-Si with OPs. The resonance line associated with interstitial Fe was only observed in the iron-contaminated sample and seems to be detectable in our experimental setup if the concentration of dissolved iron exceeds a certain threshold only. Neither the interstitial Fe nor FeB pair were observed in any sample under standard EPR measurements carried out prior each EDMR experiment. This is consistent with the fact that the maximum bulk iron concentration in our samples (1.2 $\cdot 10^{12}\:$cm$^{-3}$, measured by photodissociation of FeB pairs) is an order of magnitude lower than the sensitivity of our EPR spectrometer. The most intense features of the EMDR spectra (around $g=2.0$) have been explained in terms of P$_{\text{b0}}$ and P$_{\text{b1}}$ dangling bonds. While P$_{\text{b0}}$ dangling bonds have been observed in EDMR before, there is no consensus in the literature over whether or not the P$_{\text{b1}}$ defect is electrically active \cite{astesmans2,tdmishima1}. Our results shown in Fig.~\ref{fig4} demonstrate, however, that P$_{\text{b1}}$ is an electrically active defect and contributes to spin-dependent recombination. The recombination time (through both P$_{\text{b0}}$ and P$_{\text{b1}}$) is measured to be  $\tau = 94\:\mu$s in the precipitate-free sample (Fig.~\ref{fig7}), coinciding well with the lifetime $\tau = 1/\delta = 92\:\mu$s extracted from the y-axis intercept $\delta$ of the linear fit in Fig.~\ref{fig8}. This concurrence suggests that photo-excited electron-hole pairs recombine through dangling bonds at the surface as well as through dangling bonds at the OPs. The first recombination channel gives rise to the finite y-axis intercept, the latter to the finite capture coefficient $\gamma = 7.72\cdot 10^{-8}\:$cm$^3$s$^{-1}$. 

The results obtained in this EDMR study are consistent with those in a recent study of recombination at OPs by photoconductance measurements \cite{jdmurphy1}. This other study also found an approximately linear correlation between reciprocal lifetime and strained OP density, but in this case the intercept of the plot analogous to Fig.~\ref{fig8} was very close to the origin. The difference can be explained by the high quality surface passivation substantially reducing recombination at surface-related P$_{\text{b0}}$ and P$_{\text{b1}}$ centers. It is interesting to note that the size of the precipitate, as governed mainly by the growth time, does not seem to make a substantial contribution to the rate of recombination. The previous study \cite{jdmurphy1} tentatively suggested the size independence could be explained by recombination at the precipitate corners (see Figs. \ref{fig1} (a) to (d)). The EDMR results presented here show a linear dependence of the recombination rate due to P$_{\text{b0}}$ and P$_{\text{b1}}$ dangling bonds. This is why we speculate that P$_{\text{b0}}$ and P$_{\text{b1}}$ dangling bonds form at the corners of the strained precipitates.

\section{Summary}
In summary, we have clarified the microscopic mechanism giving rise to the EDMR effect in Cz-Si with OPs. Spin-dependent recombination of photo-excited electron-hole pairs has been identified as the predominant EDMR mechanism. We observe two coexisting defect configurations of dissolved iron (interstitial Fe and FeB) and show that both of them do contribute to spin-dependent recombination. We have demonstrated that both, the electronic recombination time and the capture coefficient can be measured with EDMR by changing the modulation frequency. Our recombination time analysis on precipitate-free and on a series of precipitate containing samples with different concentrations of OPs has shown in particular that photo-excited electron-hole pairs recombine through P$_{\text{b0}}$ and P$_{\text{b1}}$ dangling bonds formed at the sample surface and OPs. The recombination rate associated with OPs was found to increase approximately linearly with an increasing density of strained OPs with the capture coefficient $\gamma = 7.72\cdot 10^{-8}\:$cm$^3$s$^{-1}$. Further insight into the recombination process and its dynamics may be obtained from pulsed- \cite{cboehme2} and high-field EDMR \cite{vlang1} experiments, which will allow us to determine any coupling between the different centers and to study spin-dependent recombination with an enhanced spectral resolution, respectively.

\begin{acknowledgments}
The authors thank D. Gambaro, M. Cornara, and M. Olmo of MEMC Electronic Materials Inc. for performing  precipitation treatments and characterization, and V.Y. Resnik at the Institute of Rare Metals (Moscow) for performing TEM analyses. We also thank H. H\"ubl for fruitful discussions and acknowledge funding from Konrad-Adenauer-Stiftung e.V., EPSRC DTA, and Trinity College Oxford. J.D.M. is supported by the Royal Academy of Engineering, EPSRC, and St. Anne's College Oxford, and J.J.L.M. by The Royal Society, and St. John's College Oxford.
\end{acknowledgments}



\begin{thebibliography}{46}
\bibitem{footnote1}
Beside enhancing the signal-to-noise ratio, this lock-in technique also allows the Land\'{e} $g$ factor to be measured more accurately as the zero-crossing of the first derivative line is easier to determine than the position of the maximum of the EDMR absorption line.
\bibitem{footnote2}
We defined the EDMR signal amplitude $|\Delta\rho|$ as the difference between the absolute maximum and minimum of the dangling bond resonance (Fig.~\ref{fig4}). The electronic recombination time $\tau$ thus corresponds to the total recombination time of charge carriers through P$_{\text{b0}}$ and P$_{\text{b1}}$ dangling bonds. We have also measured the recombination time of the P$_{\text{b0}}$ and P$_{\text{b1}}$ center separately by extracting the different contributions to the dangling bond resonance and their individual recombination times according to Eqn. \ref{solution}. This measurement revealed the same recombination time for both, P$_{\text{b0}}$ and P$_{\text{b1}}$ dangling bonds.
\end{thebibliography}
\end{document}